\documentclass[preprint]{aastex}
\usepackage{amssymb,amsmath}
\begin{document}

\title{The 5:1 Neptune Resonance as Probed by CFEPS: Dynamics and Population}
\slugcomment{Accepted to AJ 24 Apr 2015}

\author{R. E. Pike\altaffilmark{1}}

\author{J.  J.  Kavelaars\altaffilmark{1,2}}
\author{J.  M.  Petit\altaffilmark{3}}
\author{B. J. Gladman\altaffilmark{4}}
\author{M.  Alexandersen\altaffilmark{4}}
\author{K. Volk\altaffilmark{4}}
\author{C. J. Shankman\altaffilmark{1}}

\altaffiltext{1}{Department of Physics and Astronomy, University of Victoria, Victoria, BC, Canada}
\altaffiltext{2}{National Research Council of Canada, Victoria, BC, Canada}
\altaffiltext{3}{Institut UTINAM, CNRS - Universit\'{e} de Franche Comt\'{e}, Besan\c{c}on, France}
\altaffiltext{4}{University of British Columbia, Vancouver, BC, Canada}

\begin{abstract}
The Canada-France Ecliptic Plane Survey discovered four trans-Neptunian objects with semi-major axes near the 5:1 resonance, revealing a large and previously undetected intrinsic population.
Three of these objects are currently resonant with Neptune, and the fourth is consistent with being an object that escaped the resonance at some point in the past.
The non-resonant object may be representative of a detached population that is stable at slightly lower semi-major axes than the 5:1 resonance.
We generated clones of these objects by resampling the astrometric uncertainty and examined their behavior over a 4.5 Gyr numerical simulation.
The majority of the clones of the three resonant objects ($>90\%$) spend a total of 10$^7$ years in resonance during their 4.5 Gyr integrations; most clones experience multiple periods of resonance capture.
Our dynamical integrations reveal an exchange between the 5:1 resonance, the scattering objects, and other large semi-major axis resonances, especially the 4:1, 6:1, and 7:1.
The multiple capture events and relatively short resonance lifetimes after capture suggest that these objects are captured scattering objects that stick in the 5:1 resonance.
These 5:1 resonators may be representative of a temporary population, requiring regular contributions from a source population.
We examined the dynamical characteristics (inclination, eccentricity, resonant island, libration amplitude) of the detected objects and their clones in order to provide an empirical model of the orbit structure of the 5:1 resonance.
This resonance is dynamically hot and includes primarily symmetric librators.
Given our orbit model, the intrinsic population necessary for the detection of these three objects in the 5:1 resonance is 1900$^{+3300}_{-1400}$ (95\% confidence) with H${_g}<$ 8 and $e > 0.5$.

\end{abstract} 


\section{Introduction}

The Trans-Neptunian Objects (TNOs) in the outer Solar System are the remnants of the original planetesimal disk, a record of the composition and history of the Solar System.
Many TNOs have experienced large dynamical perturbations as a result of giant planet interactions, particularly with Neptune \citep[e.g.][]{malhotra, nice, batygin2011, dawson2012}.
The dynamical history of the outer Solar System results in a complex dynamical structure in this region.
The TNOs are dynamically classified into sub-populations.
The clump of TNOs with low eccentricity, $e$, and low inclination, $i$, found between 42-47 AU is referred to as the cold classical Kuiper belt \citep{Kuiper1951}.
These objects may have undergone little dynamical perturbation.
A large number of objects, however, are currently dynamically interacting with Neptune.
These scattering objects have pericenters, $q$, between $\sim$30-37 AU, such that giant planet interactions cause changes in orbital properties (such as semi-major axis, $a$) of the TNO on timescales of 10 million years \citep{Gladman_nomenclature}.
Objects with larger $q$ are no longer interacting with Neptune and are considered detached \citep{Gladman02, Delsanti, Gladman_nomenclature}.
A significant fraction of TNOs are trapped in mean motion resonances with Neptune \citep{Gladman_nomenclature}.
The objects trapped in these mean motion resonances provide a unique diagnostic of the dynamical history of the outer Solar System \citep{cfeps_res} because resonance dynamics preserve TNOs which would otherwise scatter out of their present locations.
Mapping and cataloging these TNO populations will provide insight into the planetesimal disk's origins and evolution.

The resonance structure in the outer Solar System is extremely complex.
Significant populations of TNOs have been identified in the 3:2 resonance, at $\sim$39 AU where Pluto resides, and in the 5:1 resonance at 88 AU, as well as in many other Neptune mean motion resonances.
\cite{lykawka} find that resonance sticking occurrs out to 250 AU, which implies there may be other resonance populations, such as 6:1 resonators, still undetected.
The resonances each have different trapping efficiencies \citep{lykawka}, and if the resonances share the same source population their trapping efficiencies constrain their relative populations.
Many of the closer resonances (7:4, 5:2, 5:3, etc.) appear to contain objects with colors similar to the cold as well as the hot classical belt objects \citep{Sheppard2012} and it is unknown if this trend continues to larger $a$.
One must, however, be cautious when comparing colors of different populations of TNOs as detection biases can produce misleading results \citep{pike}.
Part of these resonant populations (of all surface types) were likely captured through a sweep up process during giant planet migration, similar to the capture of Pluto described by \cite{malhotra}.
Objects can also enter a resonance via a phenomenon referred to as `resonance sticking' \citep{duncan1997, Gomes2005, lykawka2006,lykawka}.
In this process, objects in the scattering disk population experience orbital perturbations that move them into the resonant phase space.
The objects then become temporarily ($\sim10^5$ years) or long-term ($>10^7$ years) stable in the resonance.
Temporary resonance sticking extends the lifetimes of scattering objects \citep{lykawka2006}, so these populations are dynamically linked.
The complex dynamical structure of the known resonant objects can be used to constrain the possible resonance capture mechanisms and source populations for the resonances.

The resonance structure in the outer Solar System has been explored by analytical models and dynamical simulations.
The structure of the 5:1 resonance in eccentricity and semi-major axis space is described in \cite{morbidelli1995}, demonstrating that the resonance width is greatest at $e\sim0.5$.
The n:1 resonances contain symmetric and asymmetric libration islands \citep{beauge, malhotra1996, murray-clay, cfeps_res} that determine where the TNOs come to pericenter relative to Neptune.
The n:1 resonances are the strongest resonances in the scattered disk \citep{gallardo2006}.
The Kozai mechanism \citep{kozai1, kozai2}, characterized by an exchange of eccentricity and inclination, occurs in portions of resonance phase space.
\cite{gallardo2012} find that the Kozai mechanism occurs in the n:1 resonances when the inclination is higher than a critical value which increases with semi-major axis.
This lifts the pericenters of the resonant objects.
The orbital characteristics of the intrinsic population is required for the interpretation of 5:1 resonant TNOs discovered in surveys.

Discovery biases complicate all TNO studies, but the complex structure of the 5:1 resonance makes observational studies particularly challenging.
Typical survey biases include limiting magnitudes, latitude of pointings, observation cadence, and followup methods.
Resonant populations also have a longitudinal bias because these populations come to pericenter (maximum detectability) at a specific range of longitudinal offsets relative to Neptune.
The particular location of these pericenters depends on the resonance structure; the 5:1 resonance contains both symmetric and asymmetric libration islands.
Ignoring libration, the 5:1 objects come to pericenter at approximately 180$^{\circ}$ (symmetric), 120$^{\circ}$ (leading), and 240$^{\circ}$ (trailing) ahead of Neptune; the precise locations are $e$ dependent.
As a result, a survey pointing near one of these pericenter locations would be biased toward the discovery of 5:1 TNOs from a particular island.
Pointing away from these locations can reduce the likelihood of discovery because these objects are typically only brighter than survey detection limits when they are near pericenter.
A well characterized survey with accurate pointing and depth information is needed to understand the size and substructure of the 5:1 resonance population \citep{jones}.

In this paper we examine four objects discovered near the 5:1 resonance to determine their dynamical classification, and then we use the characteristics of the 5:1 resonant TNOs to build a parametric model of the population.
Our observations and astrometry are presented in Section 2.
In Section 3 we characterize the uncertainty in our measurements and show the results of our orbital integrations of the clones of these TNOs.
We create a parametric model of the 5:1 resonance and determine a population estimate in Section 4.
Finally, Section 5 and Section 6 contain a discussion of our results and our conclusions.

\section{Observational Data}

The Canada-France Ecliptic Plane Survey (CFEPS), a characterized 803 degree$^2$ survey dedicated to detecting and tracking TNOs, discovered four objects with semi-major axes near the 5:1 Neptune resonance, at $\sim$88 AU \citep{cfeps, hilat}.
The first portion of the survey (321 degree$^2$) targeted the ecliptic plane; these fields were located within a few degrees of the ecliptic plane.
The second portion of the survey (482 degree$^2$) pointed off the ecliptic with the goal of providing constraints on the inclination distributions of TNOs by characterizing detections at higher latitudes.
The higher latitude blocks have a lower TNO discovery rate, but these fields provide a lever arm on the inclination widths of TNO populations and discover some interesting high inclination objects.
L3y02 was discovered in the ecliptic CFEPS fields (2003), and HL7c1, HL7j4, and HL8k1 were discovered in the high latitude fields (2007-2008).
All four of these objects have large inclinations (20$^{\circ}$-50$^{\circ}$), requiring the intrinsic population to be dynamically excited.
The combined CFEPS ecliptic and high latitude fields constrain the inclination distribution of the intrinsic population better than purely ecliptic surveys because high $i$ populations spend more time at high latitudes and are more difficult to discover in ecliptic surveys.

After orbital determination in CFEPS, a number of objects were flagged for extended astrometric followup because large $a$ resonant objects require longer arcs to conclusively classify the objects \citep{jones2010}.
Additional astrometry was taken over several years using Megacam on the 6.5 meter Magellan telescope in Las Campanas Observatory, Chile and the Gemini Multi-Object Spectrograph (GMOS) in imaging mode on the 8.1 meter Gemini North telescope on Mauna Kea, Hawaii, USA.
We focus here on the data from Magellan and Gemini, see Table \ref{astrom}.
Data were taken for two of the targets using Magellan Megacam in $r$ band in 2010 and 2011.
The GMOS data from 2012-2013 included observations of each of the targets for multiple nights in each of 3-4 dark runs, approximately 1-2 months apart.
The GMOS-N imaging data was taken in queue mode, which is optimal for short exposures of a few targets spread over several months.
We observed in $r$ band with a target signal to noise ratio of 10 which gives a centroid precision comparable to the catalog precision.
We took a single 7 second exposure of each target field to ensure a non-saturated image of the field stars used for astrometric calculation.
Our longer exposure times were based on the magnitude of each object, and we observed each object twice per dark run when possible.
The telescope tracked at the stellar rate of motion, which caused negligible extension in the TNO PSF due to motion ($\lesssim$1.7 arc seconds per hour).
Some elongation of the TNO was apparent in the 300 second exposures, but the PSF core was sufficiently well defined for a centroid calculation with comparable precision to the astrometric catalog.
The astrometry from Gemini and Magellan were then used to refine the orbital fits.

\begin{table}[h!]
\setlength{\tabcolsep}{4pt}
\small
\footnotesize
\caption{Astrometric Images.}
\label{astrom}
\begin{center}
\begin{tabular}{ l c  c  c  c   }
\hline\hline
Object & Date &Telescope &  Exposure Time\\\hline
L3y02 & 2013/02/05 & Gemini & 2$\times$100 sec.  \\
 & 2013/03/19 & Gemini & 2$\times$100 sec.  \\
 & 2013/04/04 & Gemini & 2$\times$100 sec.  \\
 & 2013/04/05 & Gemini & 2$\times$100 sec.  \\
\hline
HL7j4 & 2013/04/03 & Gemini & 1$\times$100 sec.  \\
 & 2013/04/04 & Gemini & 1$\times$100 sec.  \\
 & 2013/04/30 & Gemini & 1$\times$100 sec.  \\
 & 2013/05/03 & Gemini & 1$\times$100 sec.  \\
 & 2013/06/10 & Gemini & 1$\times$100 sec.  \\
 & 2013/06/15 & Gemini & 1$\times$100 sec.  \\
 & 2010/04/21 & Magellan & 1$\times$120 sec.  \\
 & 2011/05/02 & Magellan & 2$\times$120 sec.  \\
 & 2011/05/03 & Magellan & 1$\times$120 sec.  \\
\hline
HL7c1 & 2013/02/06 & Gemini & 2$\times$122 sec.  \\
 & 2013/02/06 & Gemini & 2$\times$122 sec.  \\ 
 & 2013/03/04 & Gemini & 2$\times$122 sec.  \\
 & 2013/03/19 & Gemini & 2$\times$122 sec.  \\
 & 2013/04/03 & Gemini & 2$\times$122 sec.  \\
 & 2013/04/04 & Gemini & 2$\times$122 sec.  \\
\hline
HL8k1 & 2013/04/09 & Gemini & 3$\times$300 sec.  \\
  & 2013/04/09 & Gemini & 3$\times$300 sec.  \\
  & 2013/05/03 & Gemini & 3$\times$300 sec.  \\
  & 2013/05/12 & Gemini & 3$\times$300 sec.  \\
  & 2013/06/17 & Gemini & 3$\times$300 sec.  \\
  & 2013/07/11 & Gemini & 3$\times$300 sec.  \\
  & 2013/07/15 & Gemini & 3$\times$300 sec.  \\
  & 2010/04/21 & Magellan & 1$\times$400 sec.  \\
  & 2011/05/02 & Magellan & 2$\times$400 sec.  \\
\hline
\end{tabular}
\end{center}
\end{table}

We calibrated the data prior to determining the coordinates of the TNOs.
The observations were processed and analyzed using the Image Reduction and Analysis Facility \citep[IRAF]{iraf}.
Baseline reductions as prescribed by Gemini were performed on the GMOS images.
This includes a bias subtraction, flat fielding, and combining the different amplifier chips into a single image.
For the data from Gemini and Magellan, TNOs were identified based on their predicted positions and confirmed using object motion in the case of multiple exposures.
However, the predicted positions were sufficiently accurate for the easy identification of the target even without identifying its motion with respect to the stars.
An astrometric plate solution tied to either the The Two Micron All Sky Survey \citep[2MASS]{2mass} or United States Naval Observatory \citep[USNO]{usno} catalog (depending on the availability of sources) was calculated.
We achieved an astrometric fit RMS $\lesssim$0.15 arc seconds, which limits the precision of our observations to the accuracy of the astrometric reference frame.
Astrometry was reported to the Minor Planet Center\footnote{http://www.minorplanetcenter.net/} (MPC) and used in object orbit predictions.

We performed an orbital fit \citep{bernstein} to all known astrometry of these objects, and all 4 TNOs (Trans-Neptunian Objects) were still near the 5:1 resonance.
These new astrometric data were combined with astrometry previously acquired using CFHT, earlier Magellan data, and additional resources (L3y02: University of Wisconsin-Madison, Indiana University, Yale University, and the National Optical Astronomy Observatories Telescope (WIYN), Palomar Hale, and 2.1-m at Kitt Peak; HL7c1: Palomar and WIYN; HL7j4: WIYN, Palomar, Cerro Tololo Inter-American Observatory, and Nordic Optical Telescope; HL8k1: Subaru, WIYN, and Palomar;  \cite{hilat}).
Many years of high precision astrometry is often necessary for resonance classification \citep{Gladman_nomenclature} because  resonance behavior can only be securely identified using forward numerical integration of the object's state vector.
Table \ref{obj_info} shows the results of orbital fits for the objects using \cite{bernstein}.
These objects have an orbital period of 820-830 years, so the orbital fit is based on observations of only $\sim$1\% of the orbit.
This small arc requires extremely accurate astrometry, like the measurements presented here, in order for integrations to conclusively determine resonance occupation.
Our numerical integration of clones (presented in the next section) illustrate the complexity of classifying these objects.

\begin{table}[h!]
\setlength{\tabcolsep}{2.5pt}
\small
\footnotesize
\caption{Nominal Object Orbit Fit.  The arc lengths show the years of the earliest and most recent astrometry.  The number of astrometric points, $n$, is provided as well; this large number of measurements is necessary in order to characterize the objects' orbits.  All digits shown are significant based on the barycentric orbital fit from \cite{bernstein}.  The distance at discovery is $d$.  L3y02 was measured in $g$ and $r$ band, so the $H_g$ magnitude of that object is calculated.  The other 3 objects were measured in $r$, so their approximate $H_g$ is given, assuming a $g-r=0.5$ conversion.}
\label{obj_info}
\begin{center}
\begin{tabular}{ l c  c  c  c  c  c  c c  c c c c }
\hline\hline
MPC &CFEPS &Observational & $n$ & $a$ & $e$ & $i$ & $\Omega$ & $\omega$ & T & Epoch &$d$ & H$_g$ (H$_r$)\\
ID & ID& Arc [years] & & [AU]&&[$^{\circ}$] &[$^{\circ}$] &[$^{\circ}$]  & [JD] & [JD]  & [AU] & [mag]\\\hline
2003 YQ$_{179}$&L3y02 &2003-2013& 32 & 88.41 & 0.5787 & 20.874 & 109.793 & 30.558 & 2458824.134 & 2452998.0 & 39.3 & 7.3 (6.6) \\
2007 FN$_{51}$ &HL7c1&2007-2013& 40 & 87.49 & 0.6188 & 23.237 &102.286 & 9.961 & 2445268.018 & 2454180.9 & 39.1 & $\sim$7.7 (7.2) \\
2007 LF$_{38}$ &HL7j4 &2007-2013& 51 & 87.57 & 0.5552 & 35.825 & 169.293 & 12.737 & 2439708.039 & 2454263.8 & 48.4 & $\sim$6.0 (5.5) \\
2008 JO$_{41}$ &HL8k1&2008-2013& 33 & 87.34 & 0.5431 & 48.815 &153.057 & 146.552 & 2464523.688 & 2454598.0 & 44.5 & $\sim$8.4 (7.9) \\
\hline
\end{tabular}
\end{center}
\end{table}

\section{Characterization of Discoveries}

This section describes how the astrometry and its uncertainty for each of the targets was used to determine a cloud of clones of the object  and presents the results of the forward integration of these clones.

\subsection{Estimation of Orbital Uncertainties}

All astrometric measurements have inherent uncertainties that come from a variety of sources.
Errors in the initial orbit solution and location prediction can make target identification and astrometric fits difficult.
Stellar crowding, plate boundary errors, stellar motion, and sparse catalogs can make astrometric plate solutions less accurate.
Because of these issues, systematic offsets for all observations in a single observing run are common.
When a target has clusters of astrometry separated by months without observations, a systematic offset from one observing run can significantly bias the orbital fit \citep{Gladman_nomenclature}.
Accounting for all sources of uncertainty in astrometric measurements can result in a wider range of orbital fit parameters than a simple orbital fit would suggest.

During our orbital fitting procedure, we examined all outliers in the astrometric fits to determine if the discrepant astrometry should be kept.
With years of astrometry, much of it at monthly cadences, outliers are immediately obvious in the fit residuals calculated with the routine by \cite{bernstein}.
We revisited the original data for all high residual astrometric points, and re-examined the astrometric plate solution.
For most of these observations, we calculated a new astrometric position.
For several others, we found that a good astrometric solution could not be derived, which occurred as a result of low signal on the TNO (centroiding was impossible) or an insufficient number of (unsaturated) reference stars.
We replaced or removed the astrometry in the orbital fit and repeated the check for outliers.
Our careful re-measurement of discrepant points resulted in significantly improved astrometry for all of our objects.

Systematic astrometric offsets result in poor orbital predictions for TNOs observed once or twice a year for a limited period of time \citep{Gladman_nomenclature}.
The orbit parameter estimation by \cite{Gladman_nomenclature} allows systematic offsets for some of the astrometric points and calculates the largest and smallest $a$ clones possible from the astrometry.
The larger orbit uncertainties occasionally reveal the possibility of resonance behavior for TNOs near a resonance boundary.
These targets may have high $a$ or low $a$ clones which show a resonant angle libration.
\cite{Gladman_nomenclature} noted this effect for L3y02, which at the time would not have been flagged as a potentially resonant object based on the \cite{bernstein} uncertainty range.
The \cite{Gladman_nomenclature} classification is useful for targeting possible resonant objects with small or sparsely sampled arcs for additional astrometric followup, because these objects may have systematic offsets in their astrometry that are not apparent from the fit residuals.

\begin{figure}[h!]
\begin{center}
\includegraphics[width=0.71\textwidth]{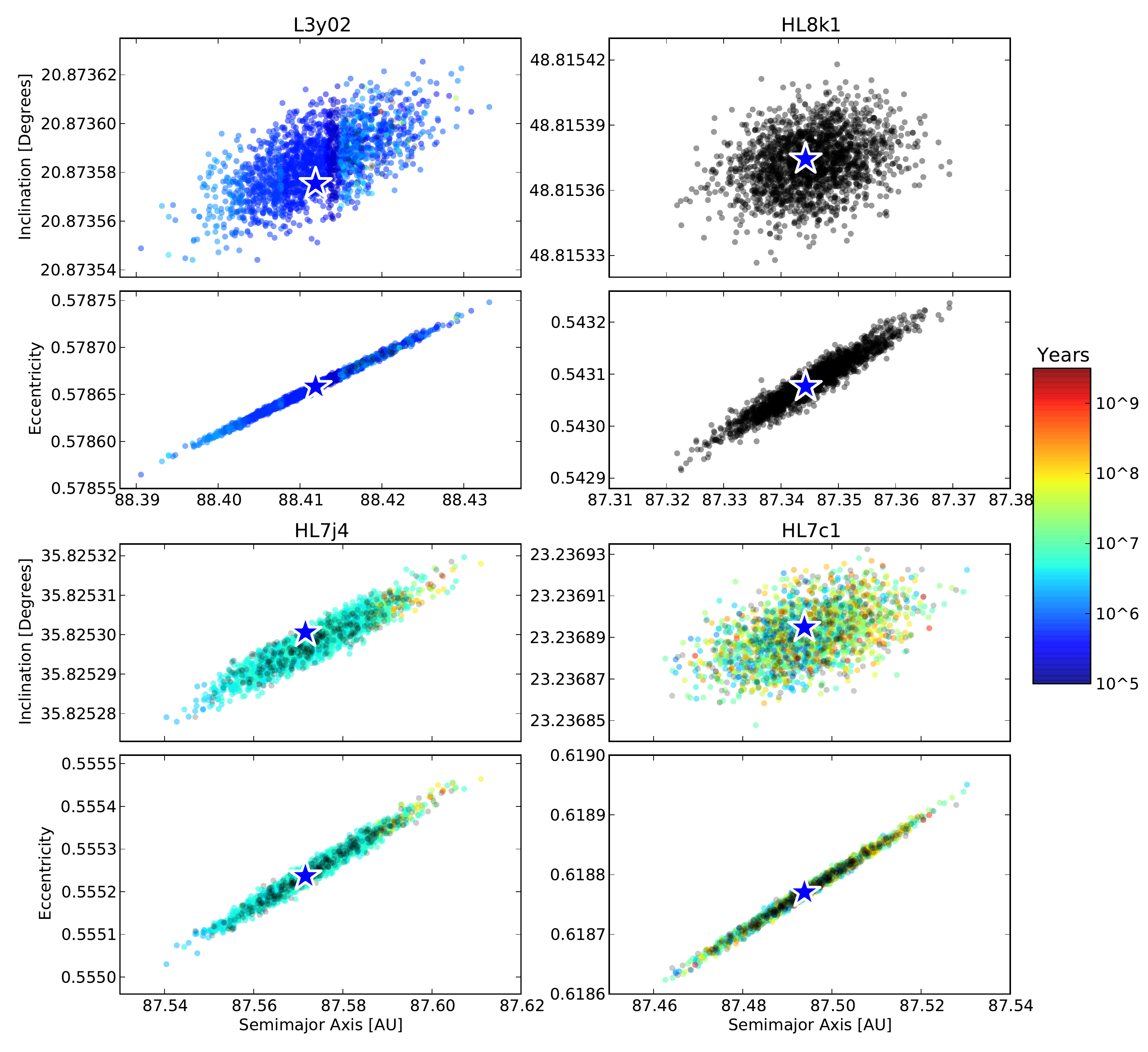}
\caption{The $\sim1500$ clones for each TNO are shown here.  The nominal best fit clone (no resampling) is shown with a blue star.  The other orbital fits were calculated by resampling the astrometry within the uncertainty and producing additional orbital fits.  This sample sufficiently explores the phase space; additional clones do not significantly alter the range of $a$, $e$, and $i$ values.  These clones provide a weighted sampling of the 1.5$\sigma$ uncertainty range of the orbit, so the median orbital behavior may be indicative of the intrinsic orbit.  The color indicates the duration of the first period of resonance occupation from the numerical integrations; if the object displays resonant behavior for $>10^7$ years the object would be classified as resonant based on \cite{Gladman_nomenclature}.  Many of the clones display multiple periods of resonance and have significantly longer total 5:1 resonance occupation.  Section 3.2 provides details on resonance diagnosis.  Black indicates a non-resonant clone.}
\label{kboclones}
\end{center}
\end{figure}

Our combination of lower than average astrometric uncertainty and better observation cadence over long arcs allows us to develop our own estimation of the uncertainty in the objects' orbital parameters.
After discrepant astrometry is removed from our results, the only sources of uncertainty in our data are the uncertainty in the position of reference stars and the centroid calculation of sources.
Our astrometry is accurate to $\sim$0.2 arc seconds.
Instead of using the method by \cite{Gladman_nomenclature}, we resampled each of our astrometric points linearly within their uncertainty of 0.2 arc seconds.
This resamples the astrometry of the objects within 1-1.5$\sigma$ uncertainty range in Right Ascension and Declination.
We calculated a new orbital fit to the resampled astrometry for the object, so we have new orbital parameters for each resampled clone.
Repeating this resampling process $\sim$1500 times provides the range of orbital parameters that are consistent with the astrometric measurements.
Using multiple orbital fits instead of resampling the uncertainty on each orbital parameter from \cite{bernstein} also allows us to preserve the interdependency of the orbital parameters.
The result was a cloud of clones of each TNO, shown in Figure \ref{kboclones}.
The behavior of the clones over time provides an exploration of the resonance behavior for orbits that are consistent with our knowledge of the intrinsic orbit of these four TNOs.

\subsection{Dynamical Integrations}

We integrated the clones of the 4 potential 5:1 objects for approximately 4.5 Gyr using SWIFT \citep{swift} in order to determine resonance occupation for each initial condition.
SWIFT integrates gravitationally interacting objects using a provided time-step, 0.5 years for this simulation.
We used the Regularized Mixed Variable Symplectic (RMVS) method which handles test particle--planet close approaches.
The position, velocity, and mass of the Sun, Jupiter, Saturn, Uranus, and Neptune were included as massive bodies, and their initial conditions were calculated for the epoch of the TNO's orbital fit.
The clones of the possible 5:1 TNOs were added as massless test particles.
We observed a variety of behaviors over the integration time, including short ($\sim10^5$ years) or long ($>10^7$ years) term resonance occupation in different 5:1 libration islands, scattering, capture into other Neptunian resonances, and stable non-resonant behavior.
Because we are interested in the region near the 5:1 resonance, test particle information was recorded only for heliocentric distances between 20 AU--150 AU.
The positions of the clones that went outside these ranges were not recorded while they were beyond the limits.
The behavior of all of the clones provides possible classifications of the TNOs found in our survey.

We determined whether an initial condition was resonant by examining the behavior of the resonant angle.
Typically, secure resonant classification requires the resonant angle to librate for 10$^7$ years from the start of integration for the best fit clone as well as the minimum and maximum $a$ clones \citep{Gladman_nomenclature}.
We classified objects as resonant if their resonant angle oscillated instead of exploring a full 360$^{\circ}$ circulation.
Following the method of \cite{trojan} for the 1:1 resonance, we used a moving window to examine the resonant angle behavior.
Because many of these objects experienced multiple periods of resonant libration, the beginning and end of each resonant period was recorded.
HL8k1 is never resonant, while L3y02, HL7j4, and HL7c1 all show periods of resonant behavior during the simulation.

\subsection{HL7j4}
\begin{figure}[h!]
\begin{center}
\includegraphics[width=.98\textwidth]{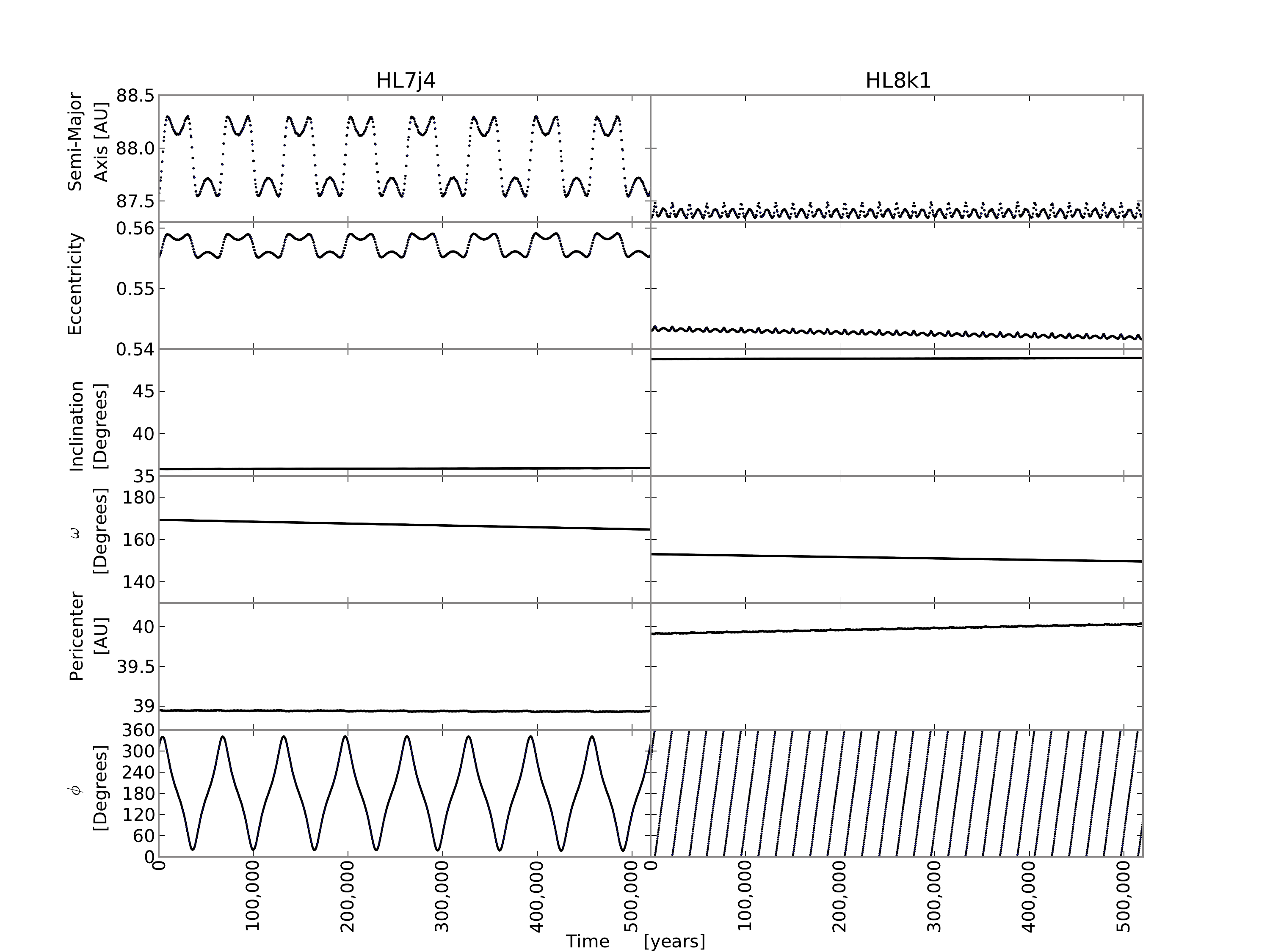}
\caption{Orbital element evolution (sampled at 300 year intervals) of the best fit clone of HL7j4 (left) and HL8k1 (right) are shown for 5$\times$10$^5$ years (representative of the behavior of both objects for the first $10^7$ years).  This clone of HL7j4 shows typical resonance behavior.  Each libration of the resonance angle corresponds to residence in both the maximum and minimum $a$ in resonance.  The majority of the clone's time is spent at these extreme $a$ values, near the resonance boundary.  The matching oscillations in $e$ result in a nearly constant $q$ for the object.  This clone of HL8k1 is not resonant (similar to all other clones of HL8k1); its resonance angle circulates.  This object appears to be in a stable position just slightly sunward of the 5:1 resonance with a slightly lower $a$ than the resonance border.}
\label{resonance}
\end{center}
\end{figure}

Based on our simulations, HL7j4 is in the 5:1 resonance.
The behavior of the best fit clone over the first 500,000 years of the simulation is shown in Figure \ref{resonance}.
This symmetric oscillation was observed for all of the clones for a minimum of 2$\times$10$^6$ years, so we classify this object as a symmetric 5:1 librator.
The median duration of the first period of resonance is 4.9$\times10^6$ years.
See Figure \ref{kboclones} for a plot of resonance duration.
The higher occurrence of longer lived clones in the large $a$, $e$, and $i$ corner suggest that this region may be more likely to contain the `true' orbit of the TNO.
When the resonant periods for each clone are summed, more than 99\% of the clones are resonant for at least a total of 10$^7$ years of the entire simulation, and 80\% are resonant for 10$^9$ years.
Of the clones that leave the 5:1 resonance, several are captured into other distant resonances, including the 6:1 and 7:1 resonance.
Of our four objects, HL7j4 has the highest fraction of resonant clones that are also in the Kozai resonance \citep{kozai1,kozai2}, which can lead to larger variations in the test particles' eccentricities and inclinations.
The clones of HL7j4 display the most stable resonant behavior of our test particles, and the `true' TNO may be on an orbit in a portion of the phase space stable for $\sim$Gyr.

\subsection{HL7c1}
HL7c1 is also a 5:1 resonator.
The best fit clone is resonant for the first 6$\times10^7$ years, before exiting the resonance.
The clone experiences two additional short resonance captures ($\sim10^5$ years) before scattering outward.
The best fit clone passes beyond 150 AU at 3.2$\times10^8$ years, then re-enters the region of interest at 7.1$\times10^8$ years.
It moves beyond 150 AU again briefly before scattering inward at 9.5$\times10^8$ years where the simulation output ends.
All clones of this object begin in symmetric 5:1 resonance, and remain in symmetric resonance for a minimum of $10^6$ years.
The duration of the first resonance occupation for these clones is shown in Figure \ref{kboclones}.
The seemingly random distribution of resonance occupation times suggests that no particular $aei$ region is favored.
The median duration of the first period of resonance is 1.6$\times10^7$ years.
Approximately 90\% of the clones of this object are resonant for a total of at least 10$^7$ years.
However, only 30\% are resonant for a total of 10$^8$ years and $<$10\% for a total of 10$^9$ years.
After the clones leave the resonance, the majority move into scattering orbits and some become detached objects.
Some of these scattering clones are captured into other resonances and some are ejected from the region of our SWIFT output (20AU-150AU).
HL7c1 is currently a 5:1 resonator.

\subsection{L3y02}
L3y02 is also classified as resonant.
The best fit clone begins with $5\times10^4$ years of resonance, exits the resonance, then re-enters the resonance at $2\times10^6$ years.
Similar behavior is observed for the other clones; there are multiple periods of resonance broken by non-resonant behavior.
All of these clones begin their first resonance as symmetric librators.
The median duration of the first period of resonance is 6.0$\times10^5$ years.
Most of the resonance periods are of short duration; the length of resonance for each clone is shown in Figure \ref{kboclones}.
Over 95\% of the clones are resonant for a total of at least 10$^7$ years during the simulation, 80\% for 10$^8$ years, and 50\% for 10$^9$ years during the simulations.
When leaving the resonance, these clones preferentially exit to low $a$, just sunward of the 5:1 resonance.
Many clones remain there, becoming detached objects while the others enter the scattering population, some of which are captured into other resonances.

\subsection{HL8k1}
HL8k1 is not in the 5:1 Neptune resonance, in spite of being extremely close to the resonance boundary.
All of its clones experience small periodic oscillations in $a$ and $e$, shown in Figure \ref{resonance}, but not with the characteristic periodic pattern seen in the resonant object integrations.
The resonant angle shows only a slow circulation instead of a bounded oscillation, so HL8k1 is classified as a detached object according to the \cite{Gladman_nomenclature} classification scheme.
HL8k1 is not resonant, but it may have been associated with the 5:1 resonance at some point in the past.

\subsection{Resonance Characteristics}
The resonant angle behavior in Figure \ref{resonance} for HL7j4 is representative of a typical symmetric libration behavior.
Most of the clones of our three 5:1 objects are primarily symmetric librators, although a small number of clones transition into periods of asymmetric libration.
The resonance angle $\phi$ librates with a high amplitude centered around 180$^{\circ}$.
This large amplitude symmetric libration was typical for $\sim90$\% of the resonant clones; the smaller amplitude asymmetric libration was far less common and typically occurred between periods of large amplitude symmetric libration.
The intrinsic asymmetric fraction of the resonance could be as high as $\sim$30\% if the fraction is similar to the n:1 resonance symmetric fractions from \cite{cfeps_res}.
Our smaller observed fraction may be the result of the longitudinal biases of the discovery surveys.
Understanding the distribution of the libration amplitudes and libration centers will be critical to understanding the 5:1 resonance capture process.

\begin{figure}[h!]
\begin{center}
\includegraphics[width=1.1\textwidth]{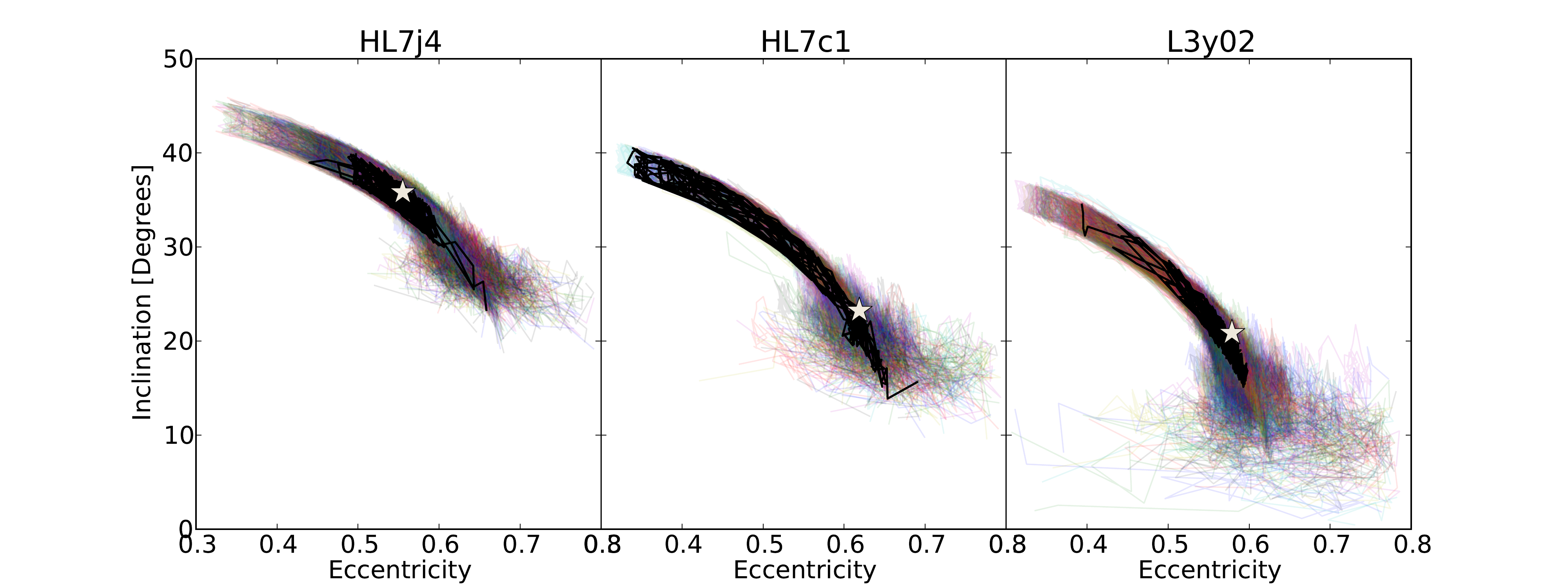}
\caption{Each colored line shows the evolution of a single clone over time (up to 4.5 Gyr) in eccentricity and inclination.  A typical clone in Kozai is bold in each plot.  The plot includes all clones with a mean $87<a<89$, which limits the plot to primarily resonant TNOs.  The current best fit orbital parameters for each object are marked with a star.  The Kozai mechanism in the 5:1 resonance is apparent at the upper left of each plot; the resonant objects that evolve to low eccentricity experience a simultaneous increase in their inclination.  The Kozai evolution track is obvious in this plot, however, these clones are only $\sim$5\% of the test particles.}
\label{i_vs_e}
\end{center}
\end{figure}

\begin{figure}[h!]
\begin{center}
\includegraphics[width=0.8\textwidth]{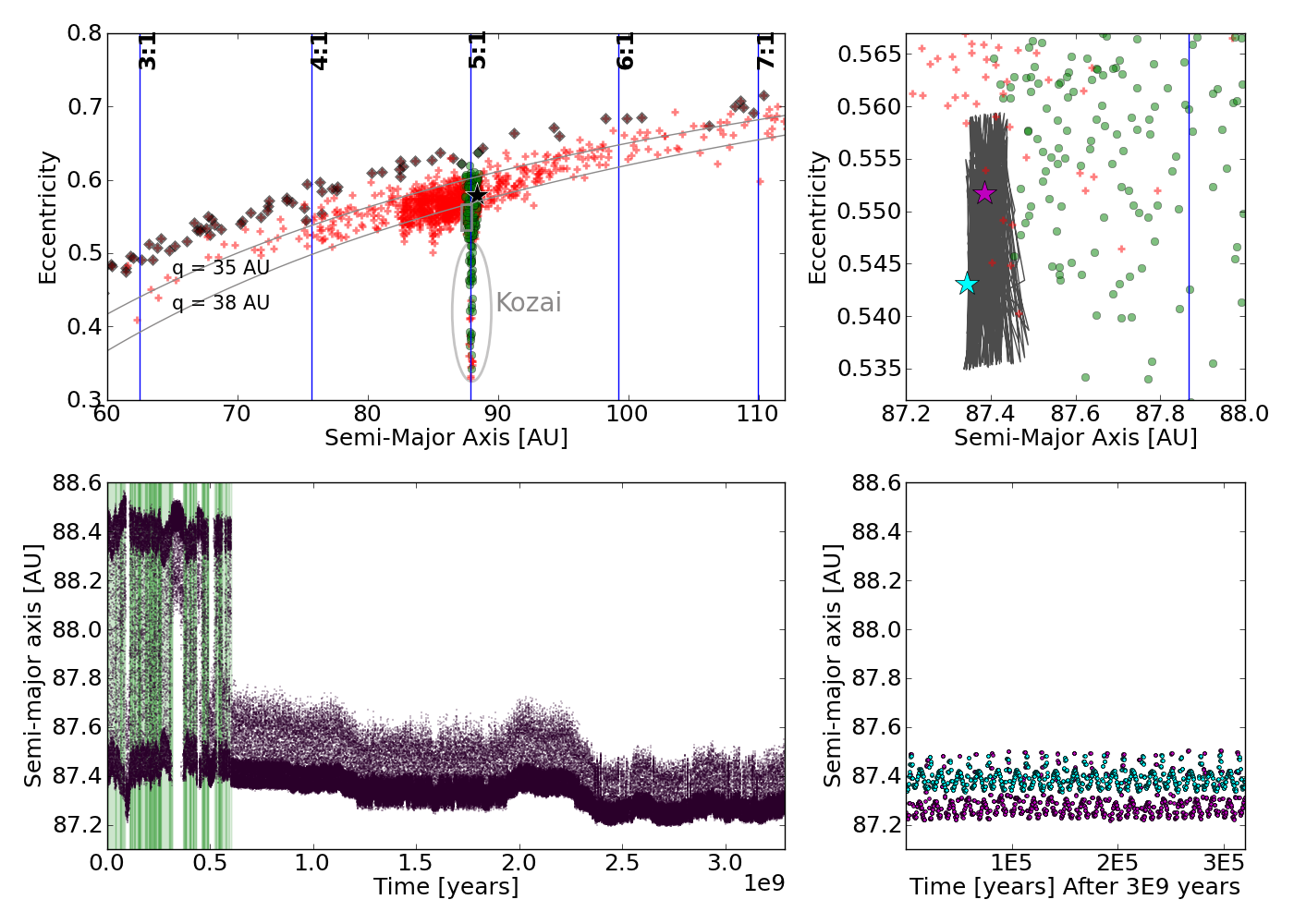}
\caption{
Upper Left: This end state plot plot shows all clones of the resonant object L3y02.  The clones which are 5:1 resonators at that time are marked in green circles, non-resonant clones are red `+', and the last recorded position of `lost' objects are marked with black diamonds.  The black star indicates the initial conditions of L3y02.  The grey arcs mark the `$q$=35 AU' and `$q$=38 AU' lines.  The clones that have traveled down the 5:1 resonance (decreasing eccentricity) create a population beyond our detection limits as a result of Kozai cycling.  This snapshot also contains escaped clones which have been captured into the 3:1 and 7:1 resonance.  The overabundance of objects just inside the 5:1 resonance is a pseudo-stable population, likely produced by chaotic diffusion away from the resonance border, of which HL8k1 may be a member.  Upper Right: The zoomed in region from the grey rectangle in the left plot shows the non-resonant objects just outside the 5:1 resonance boundary.  The grey line traces the path of the nominal non-resonant clone of HL8k1 (cyan star) for 3$\times10^9$ years.  The magenta star is an end state clone of L3y02 that has diffused out of the resonance.  Lower Left: The evolutionary history in $a$ of the magenta resonance diffusion L3y02 clone.  Resonant periods are shaded green.  Lower Right: A zoomed plot of $a$ for this non-resonant L3y02 clone beginning at 3$\times10^9$ years.  The $a$ evolution is similar to HL8k1 (cyan).}
\label{L3y02_clones}
\end{center}
\end{figure}

The detected 5:1 TNOs all have large eccentricity ($>0.54$), but $\sim$5\% of their resonant clones were found to undergo a decrease in eccentricity as a result of resonant dynamics.
The eccentricity of resonant clones ranged from approximately 0.65 to 0.30, with the majority of clones having values between 0.5-0.65.
After 10$^9$ years 3\% of resonant clones had cycled to an $e$ below 0.5, and after 2x10$^9$ years 5\% of the remaining resonant clones had cycled to an $e<0.5$.
This cycling is visible in Figure \ref{i_vs_e} and Figure \ref{L3y02_clones}.
This suggests that a Kozai mechanism \citep{kozai1, kozai2}, where there is an exchange between eccentricity and inclination, is acting on the clones.
The evolution of $e$ and $i$ results in objects with inclinations up to 45$^{\circ}$.
These objects remained primarily resonant, with low $e$ and larger $q$ than the detected objects in the 5:1 resonance.
The greater stability of low $e$ objects in the 5:1 resonance, combined with the supply of this population from large $e$ objects, implies that there is likely a population in the 5:1 resonance stored at low $e$.
Many of the 5:1 resonators that are also in Kozai resonance will cycle back to a larger $e$ over time, so the fraction of objects at low $e$ remains relatively constant after 10$^9$ years.
Current survey depths make it nearly impossible to detect all but the brightest low $e$ 5:1 objects; for CFEPS with limiting magnitude of $g\sim$23.5-24.4 \citep{cfeps}, a 5:1 TNO with $e=0.3$ (q$\sim$61 AU) would have to have an absolute magnitude $H_g\sim$5.6-6.5 in order to be detected at perihelion.
We did not detect any low $e$ objects, so to avoid complication from this possibly interesting component we ignore this $\sim$5\% component of low $e$ objects that may have evolved from larger $e$ objects when we calculate our population estimate in Section 4.

The clones of the resonant objects that left the resonance with $e$ values such that they do not immediately scatter off Neptune were preferentially deposited or retained at $a$ and $e$ values similar to the current orbit of HL8k1, the non-resonant object.
This overabundance of clones is shown in Figure \ref{L3y02_clones}.
These objects are found along the resonance borders and at a slightly smaller $a$ and $e$ than the 5:1 resonance.
HL8k1 exhibits stable behavior over the age of the Solar System.
Figure \ref{resonance} shows the first 5$\times10^5$ years of HL8k1's evolution, and the behavior is representative of the entire 4.5 Gyr history of all clones of HL8k1.
The clones experience regular, but small, changes in $a$ and $e$ as well as slow evolution of the resonance angle.
None of the clones of HL8k1 are captured by the resonance during their 4.5 Gyr integrations, which reflects the low probability of reentering the resonance.
This object could have been emplaced in its current location during Neptune's migration {\citep{gomes_ssbn} or diffused out of the resonance without the aid of planetary migration, which was not included in our simulations.
We use `resonant diffusion object' to refer to a TNO that appears to have leaked out of a resonance and entered a pseudo-stable region of phase space.
The near-resonant object, HL8k1, was likely dynamically associated with the 5:1 resonance and may be representative of a large number of resonant diffusion objects.

\section{5:1 Population Model}

Because of the many biases against detection, the discovery of three distant 5:1 resonant TNOs in the CFEPS survey clearly indicates that the 5:1 Neptune resonance is well populated.
We created a model population of objects based on the known characteristics of the real objects, the stability structure of the 5:1 resonance, and the parametric distributions of other known resonant populations.
The sensitivity and pointings of the \cite{alexandersen} and CFEPS surveys are known \citep{cfeps,hilat}.
We combined our model of the 5:1 resonance with the discovery biases for real objects in order to estimate the size of the intrinsic 5:1 population.

\subsection{Parametric Model of the Resonance}

Our model of the 5:1 resonance is based on the regions of dynamical stability.
The structure of the n:1 resonances is complex, due to the presence of multiple libration islands.
The symmetric resonance objects come to pericenter 180$^{\circ}$ from Neptune (plus or minus their libration amplitude which can approach 180$^{\circ}$), and the objects in the two asymmetric resonant islands come to pericenter on each side, 120$^{\circ}$ from Neptune (plus or minus a smaller libration amplitude).
These complexities make visualization difficult, but a simplified model of the 5:1 resonance is presented in \cite{cfeps_res}.
Here we use a similar parameterization of the resonance structure.

The semi-major axis range of a resonance is restricted by the resonance width.
The exact width depends on the other orbital parameters (such as inclination and eccentricity), but for the purposes of detectability it is sufficient to establish a model with a single range of $a$ values.
While the density of points at the $a$ resonance edges in the left plot in Figure \ref{resonance} make it clear that the clones spend the majority of their time near the resonance edges, the resonance is narrow enough that this complication does not have a significant impact on the detectability of model objects.
We model the $a$ range as a simple uniform distribution between 87.9 and 88.9 AU.

The objects in the 5:1 resonance appear to have a very excited inclination distribution.
The detected objects all have high inclinations, even L3y02, which was discovered in the ecliptic block of CFEPS.
The ecliptic blocks of CFEPS were sufficiently deep that 5:1 objects with low $i$ would have been detectable if they were present.
The detection of a high inclination object near the ecliptic, where high $i$ objects spend only a small fraction of their orbits \citep{cfeps_res}, as well as two object discoveries in the high latitude blocks indicates that this population is dynamically excited.

Based on our model analysis, we find that the 5:1 resonance probably has a wider inclination distribution than the Plutinos.
We implemented an inclination distribution parameterization from \cite{brown2001} which is acceptable for other studied resonances \citep{cfeps_res}.
\begin{equation}
  P(i) \propto \sin(i)\times \exp\left(\dfrac{-i^2}{2\sigma_i^2}\right)
\end{equation} 
We drew inclinations randomly from this inclination distribution for our 5:1 population model, using 5$^{\circ}<\sigma_i<65^{\circ}$ in increments of 1$^{\circ}$.
10,000 simulated detections were produced using the CFEPS survey simulator, and then the inclination of these simulated detections was compared to the three measured object detections.
We used the Anderson-Darling statistical test with a bootstrapped sample from the simulations to determine the likelihood of reproducing the three measured inclinations, shown in Figure \ref{inclin}.
HL8k1 has the largest inclination of the discovered objects, so if the fourth object were included, the inclination width would be higher.
The most acceptable inclination width, $\sigma_i$, for the three 5:1 objects was 22$^{\circ}$.
We reject inclination widths less than 15$^{\circ}$ and more than 43$^{\circ}$ at 95\% confidence.
The published inclination widths for the Plutinos are 14$^{\circ}$$^{+8}_{-4}$ \citep[95\% confidence limits]{alexandersen}, and 16$^{\circ}$$^{+8}_{-4}$ \citep[95\% confidence limits]{cfeps_res}, and 10.7$^{\circ}$$^{+2.0}_{-2.3}$ \citep[68\% confidence limits]{gulbis}.
We find that the 5:1 resonance likely has a wider inclination distribution than the Plutinos and other hot TNO populations, however we cannot rule out some similar inclination widths at 95\% confidence.

It is possible that a significant amount of objects in the n:1 resonances are resonance sticking from the scattering object population, either currently or in a primordial scattering event.
Resonance sticking objects may not include a distribution that extends to low inclinations, in which case the \cite{brown2001} distribution may overestimate the size of the resonance population.
Our sample of 3 objects is too small to rule out a low $i$ tail, but simulations of planetary migration resulting in resonance sticking show an inclination distribution not centered at zero \citep{Gomes2005}.
Our clones show a similar evolution to the particles \cite{Gomes2005}, including the influence of the Kozai mechanism on the inclination distribution.
We explore a toy model of the inclination distribution based on the $i$ distribution presented in Figure 3 of \cite{Gomes2005}; a Gaussian peak center, $\mu$, at 35$^{\circ}$ instead of 0$^{\circ}$ with a width $\sigma_i$ of 7$^{\circ}$.
This conservative parameterization removes the undetected low $i$ tail present in the \cite{brown2001} model.

\begin{equation}
  P(i) \propto \sin(i)\times \exp\left(\dfrac{-(i-\mu)^2}{2\sigma_i^2}\right)
\end{equation} 

 \begin{figure}[h!]
\begin{center}
\includegraphics[width=1.1\textwidth]{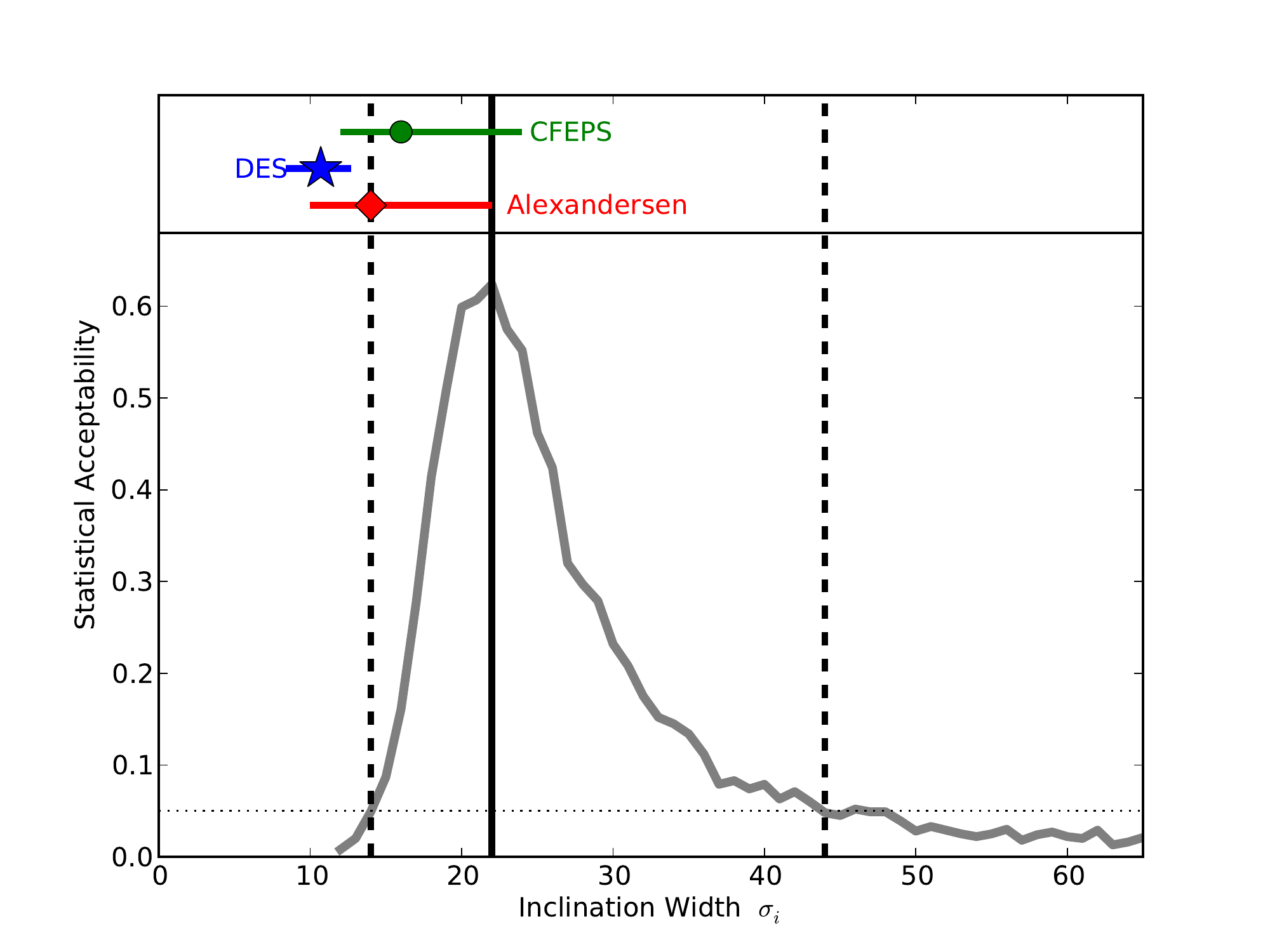}
\caption{The acceptability of different inclination widths for the 5:1 resonance is shown for inclination widths from 5$^{\circ}$-65$^{\circ}$.  The preferred inclination width is 22$^{\circ}$ (black solid line).  The 95\% limits are 14$^{\circ}<\sigma_i<44$ (black dashed lines).  The Plutino inclination width is shown for comparison from \cite{alexandersen} (red diamond, 95\% confidence limits), \cite{cfeps_res} (green circle, 95\% confidence limits), and \cite{gulbis} (blue star, 68\% confidence limits).}
\label{inclin}
\end{center}
\end{figure}

The apparent eccentricity distribution of a population at large semi-major axis is strongly biased.
Objects from the 5:1 resonance are typically only sufficiently bright for detection near pericenter, and if their pericenter is too large they may not be detectable even at pericenter.
For this reason, we did not assume an eccentricity range for our model population based on another resonant TNO population but instead used an eccentricity range based on the range in our real detections and the bulk of the integrated clones (see Figure \ref{L3y02_clones}).
We used a uniform eccentricity distribution from 0.50 to 0.62, or pericenters of $\sim$30 to 45 AU.
The known objects have eccentricities in the range 0.55 to 0.62.
While there may be a component of the 5:1 resonance with lower eccentricity due to the Kozai mechanism (suggested at the 5\% level by our simulation results) or some other mechanism not present in our simulations, we present a minimum population estimate by not extending the model to eccentricities significantly lower than our detected TNOs.

Our cloned objects explore much of the resonance space, so we used the resonant libration characteristics of the clones to constrain the libration centers and amplitudes in our parametric model.
The population model assumes 90\% symmetric and 10\% asymmetric resonators.
We did not detect any asymmetric resonators and do not place strong constraints on their population fraction.
We also calculated the range of libration amplitudes explored by the 5:1 resonators and included that in our orbital distribution model.
The symmetric objects had their expected libration centers at 180$^{\circ}$ from Neptune, and an amplitude of 175$^{\circ}$$^{+3}_{-5}$.
The trailing island was centered at 240$^{\circ}$ with an amplitude of 70$^{\circ}$$^{+20}_{-40}$, and the leading island was centered at 120$^{\circ}$ with an amplitude of 70$^{\circ}$$^{+20}_{-40}$.
We calculated positions for each model object by drawing from a triangular distribution of the libration amplitude encompassing the libration amplitude range reported.
An angle was randomly selected from within the object's libration range, instead of sinusoidally weighting the objects position.
This gives a closer approximation to the sawtooth pattern of the resonant angle seen in Figure \ref{resonance}.
This angle was added to its libration center to give the current resonant angle, $\phi_{51}$, for the model object.
As described above, the resonance characteristics of the integrated clones were used to generate a representative sample of resonance angles.

The three additional orbital parameters (argument of pericenter, mean anomaly, and longitude of the ascending node) must fulfill the resonance condition.
A detailed discussion of the resonance angles can be found in \cite{murray-clay}.
We selected the mean anomaly ($\mathcal{M}$) and the node ($\Omega$) randomly, and calculated the necessary argument of pericenter ($\omega$) using the resonant angle, $\phi_{51}$.  
This condition required Neptune's position, $\lambda_N$ = $\omega_N + \Omega_N + \mathcal{M}_N$, at the selected epoch.
The resonance condition is:

\begin{equation}
  \phi_{51} = 5\lambda -\lambda_N - 4\varpi
\end{equation} 

where $\varpi$ = $\omega + \Omega$.
This means that $\omega$ was calculated via:

\begin{equation}
  \omega = \phi_{51} - \Omega - 5M + \lambda_N  .
\end{equation}

An absolute $H$ magnitude distribution is needed in order to estimate the number of objects in the resonance.
This distribution determines the number of objects in the population of each size.
Our three 5:1 resonant objects have $H_g$ of 6.0, 7.3, and 7.7 (see Table \ref{obj_info}).
Our population estimate only extends to $H_g = 8$, providing a more accurate population estimate for this sparsely sampled resonance than extrapolating to smaller sizes.
$H_g = 8$ corresponds to an approximate diameter of 170 km \citep{cfeps}.
A single power law with a slope of $\alpha=0.8$ has been shown to be appropriate for the scattering objects to $H_g\sim$9 \citep{shankman}.
A similar result for $H_B<7.7$ was reported by \cite{fraser}, who found the `hot' population had a bright end slope of $\alpha=0.87$.
We expect our objects to be similar to scattering objects, so we used a single power law with a slope of $\alpha=0.8$ for our size distribution, and only estimate the population for $H_g < 8$.
Because we only extended our population estimate over the range of our detected objects, a slightly steeper slope, as in \cite{fraser}, does not significantly impact our population estimate.

\subsection{Population Estimate}

With the assumptions described in the previous section, we made a population model for the 5:1 resonant objects.
We used inclination width $\sigma$ = 22$^{\circ}$, the eccentricity range of 0.50-0.62, and symmetric fractions and libration amplitudes consistent with the clones, discussed above.
This results in a minimum model-based population estimate for the 5:1 resonance.

We estimated the size of the underlying population of the resonance by using a survey simulator \citep{jones, kavelaars2009}.
The pointings and depths of each survey field are provided to the simulator from the CFEPS ecliptic and high latitude blocks and the \cite{alexandersen} survey.
Analysis similar to \cite{kavelaars2009} showed consistency between the real object detection and the model objects, once biased by the survey simulator.
The survey simulator works by testing the detectability of each model object against the survey fields.
Synthetic objects were drawn from the model until 3 tracked TNOs were `discovered' by the survey simulator; the total number of objects drawn gives a possible population size.
This process was repeated 6,000 times, and a range of population estimates were computed, shown in Figure \ref{pop_est}.

 \begin{figure}[h!]
\begin{center}
\includegraphics[width=1.1\textwidth]{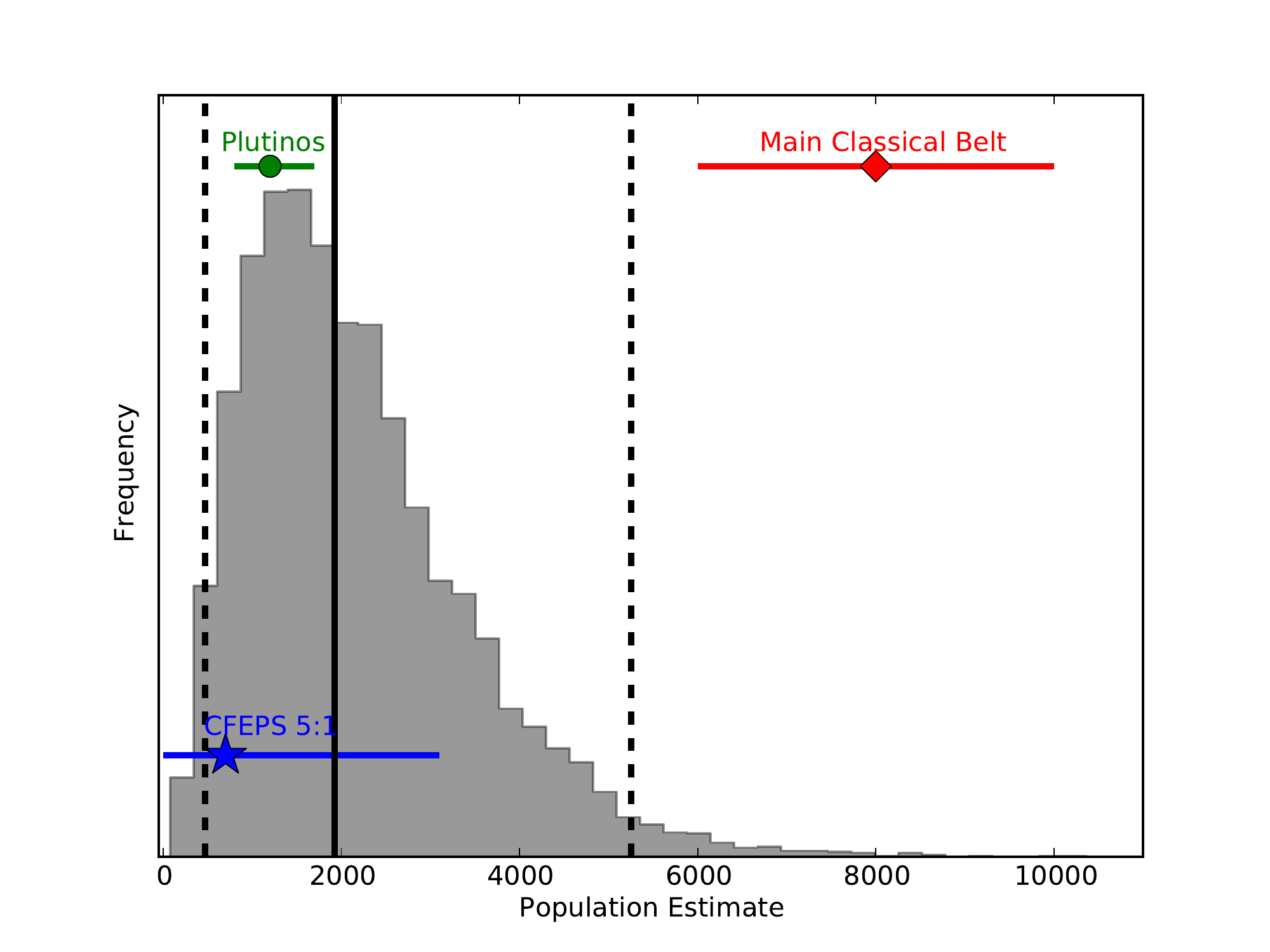}
\caption{The histogram shows the number of intrinsic objects with $H_g < 8$ necessary in the 5:1 resonance in order for the surveys to have detected 3 objects.  The wide range of acceptable values is due to the low number of detections, but the median population prediction is 1900 TNOs in the 5:1 resonance brighter than H${_g}<$ 8.  The median value is shown by the bold black line, and the 95\% confidence limits are shown in black dashed lines.  The blue star shows the \cite{cfeps_res} 5:1 population estimate.  For comparison, the population estimates (and 95\% confidence ranges) for the Plutinos (green circle) from \cite{cfeps_res} and Main Classical belt (red diamond) from \cite{cfeps} are shown.}
\label{pop_est}
\end{center}
\end{figure}

We find that the 5:1 resonance contains 1900$^{+3300}_{-1400}$ TNOs with H${_g}<$ 8 and $e > 0.5$.
This estimate is approximately the size of the Plutino population \citep{cfeps_res} and possibly larger, shown in Figure \ref{pop_est}.
Our result is consistent with the 5:1 population from \cite{cfeps_res} based on a single detection; their population model assumed a narrower inclination distribution but extended the model to $0.35 <e <0.65$.
For our toy model with the inclination distribution drawn from a Gaussian centered at 35$^{\circ}$, we find 2400$^{+4000}_{-1800}$ objects in the 5:1 resonance with $H_g<8$ and $e>0.5$.
We tested some additional model parameters to determine the extent of the model dependence of the population estimate.
Changing the symmetric fraction from 10\% to 30\% resulted in a population estimate within 1\% of our nominal model.
Using the inclination distribution extremes (14$^{\circ}$ and 44$^{\circ}$) gave population estimates 5\% larger and 17\% smaller, respectively.
This outer resonance represents a significant reservoir of TNOs, regardless of the inclination distribution of the population.

\section{Discussion}

We have found three objects in the 5:1 resonance which appear to be resonance sticking objects, indicating the existence of a large population of temporary and long term captures.
In our integrations, we found that a significant fraction of the clones escaped the 5:1 resonance into the detached or scattering population on timescales of 10$^8$ to 10$^9$ years.
Some of these objects were ejected, either inward or outward, but many remain non-resonant with semi-major axis values near the 5:1 resonance.
We noted captures into the 6:1 and 7:1 resonances, as well as some additional resonances.
We also identified cases where clones left the 5:1 resonance, entered a different resonance, escaped that resonance, and were recaptured into the 5:1 resonance.
The instability in these resonances implies a regular exchange between the scattering objects and the outer resonances.
A reliable estimate of capture efficiency from the scattering objects combined with stability lifetimes would determine the plausibility of different population mechanisms: a steady state population resupplied by the currently scattering objects, remnant long term captures from a past scattering event, or a non-scattering capture source.
We suggest that the scattering objects are the `source' of a resonance sticking 5:1 populations, and the nearby resonances with efficient resonance sticking (such as the 6:1 and 7:1 resonances) should also have large resonance sticking populations.

The detached object HL8k1, located just interior to the 5:1 resonance, may be a resonance diffusion object.
This object is stable in its current location for the age of the Solar System.
In the integrations of the clones of the resonant objects, we see an over-density of objects exiting the 5:1 and remaining near the resonance borders and just sunward of the 5:1 resonance.
Some of these clones remain in this location for the rest of the simulation, so a diffusion out of the resonance could populate the region around HL8k1 where objects are stable.
The detection biases against an object like HL8k1 are similar to those against the resonators; a first order population estimate for HL8k1-type objects is $\sim1/3$ the 5:1 resonant population.
Because of the extreme inclination of this object, we prefer the resonant population model with inclination not centered on the ecliptic for the comparison 5:1 population model.
HL8k1 suggests the existence of a resonance diffusion population in a stable region of phase space and may be representative of a large, detached resonant diffusion population; in fact HL8k1 exhibits similar behavior to 2004 XR$_{190}$ \citep[Buffy,][]{Allen06}, with a semi-major axis just sunward of the 8:3 resonance.

In our model of the 5:1 resonance, we assume only eccentricities that are stable and in the range of those detected by current surveys.
This limits our eccentricity range to 0.50-0.62.
We find that Kozai cycling can result in resonant objects with eccentricities of 0.3-0.5 and high inclinations.
We do not include these objects in the model because they have very low detectability and, based on the clone integrations, likely represent $\sim$5\% of the population.
Because the surveys have low sensitivity to low $e$ objects, we cannot rule out a low $e$ population unrelated to Kozai, with a hot or cold $i$ distribution.
If objects with both low $e$ ($<0.5$) and low $i$ ($<15^{\circ}$) were discovered in the 5:1, such objects might not be dynamically linked to the known 5:1 resonators and could require a different source population or capture mechanism.

We find that the 5:1 resonance is well populated, with 1900$^{+3300}_{-1400}$ TNOs with H${_g}<$ 8.
This supports the tentative prediction in \cite{cfeps_res} that the 5:1 resonance may be the most populous of all the known resonances.
\cite{alexandersen}, using the CFEPS ecliptic blocks and their additional survey, found a 4:1 population estimate of 230$^{+900}_{-220}$ and a 3:1 population estimate of 270$^{+360}_{-180}$ for H${_g}<$ 8, significantly less populated than the 5:1.
The 5:1 population estimate is larger than the CFEPS estimate of the Plutinos, which have the largest number of detected objects of any resonance.
The 5:1 resonance is twice as far from the Sun as the 3:2 resonance, so a different population mechanism and source population are likely.
The range of stable inclinations and eccentricities are larger in the 5:1, and it would appear, naively, that the phase space of the 5:1 resonance is significantly larger than the 3:2.
A combination of increased phase space and resonance sticking lifetimes could result in this large 5:1 population.

The only other known TNO populations at semi-major axis near the 5:1 are other possible resonances, scattering objects, and some detached TNOs.
\cite{shankman} find that there are 10$^4$ scattering objects with H${_g}<$ 8 out to $a<1,000$ AU.
This means that the 5:1 resonance is $\sim$20\% the size of the total scattering object population.
Based on the semi-major axis distribution of the model used by \citep{shankman} and the total population estimate therein, we find that the scattering population with H${_g}<$ 8 between 85-95 AU is approximately 220 objects.
The scattering objects have a density of $\sim22$ objects per AU, so the 5:1 resonance, with a width of $\sim$2 AU containing $\sim2300$ objects represents a significant over-density in this semi-major axis range.

\section{Conclusions}

Based on extensive tracking of four objects discovered by CFEPS, we have determined that three objects are in the 5:1 Neptune resonance.
We have performed orbital integrations of these objects over the age of the Solar System, and it appears that these objects are likely temporary captures, stable for $\sim10^8$ years.
These resonance sticking events can result in short or long term resonance.
The fourth object is not resonant, but is in a stable location for the age of the Solar System, and its location relative to the 5:1 resonance suggests it is a resonance diffusion object.

The characteristics of the three resonant objects and the dynamical integrations were used to create a parametric model.
We selected an inclination width, eccentricity and semi-major axis range, and the $H$ magnitude distribution of the scattering objects.
The inclination width that reproduced the inclinations of the three detected TNOs was larger than the Plutinos, suggesting a significantly dynamically hotter population.
All of the resonant objects begin in the symmetric resonance, and some clones transition into the asymmetric islands.
Approximately 5\% of the simulated objects are affected by the Kozai mechanism which pushes some objects to low $e$ and high $i$, but we exclude this small fraction of objects from our model for a population estimate.

We used a survey simulator to apply the survey biases to our model, and found that the resulting model detections resembled our real TNOs.
The necessary size of the model in order to produce three detections gives a population estimate for the 5:1 resonance of 1900$^{+3300}_{-1400}$ with H${_g}<$ 8 and $e>0.5$.
This significant population in the outer Solar System has implications for Solar System evolution as well as implications for other distant resonances because the 5:1 resonance appears to be scattering objects sticking in the 5:1 resonance.

\acknowledgements
Based on observations obtained at the Gemini Observatory processed using the Gemini IRAF package, which is operated by the Association of Universities for Research in Astronomy, Inc., under a cooperative agreement with the NSF on behalf of the Gemini partnership: the National Science Foundation (United States), the National Research Council (Canada), CONICYT (Chile), the Australian Research Council (Australia), MinistŽrio da Cincia, Tecnologia e Inova‹o (Brazil) and Ministerio de Ciencia, Tecnolog'a e Innovaci—n Productiva (Argentina).

This research used the facilities of the Canadian Astronomy Data Centre operated by the National Research Council of Canada with the support of the Canadian Space Agency.

Based on observations obtained with MegaPrime/MegaCam, a joint project of CFHT and CEA/IRFU, at the Canada-France-Hawaii Telescope (CFHT) which is operated by the National Research Council (NRC) of Canada, the Institut National des Science de l'Univers of the Centre National de la Recherche Scientifique (CNRS) of France, and the University of Hawaii. 

Research supported by the Canadian National Science and Engineering Research Council (NSERC) Discovery Grant program.

\end{document}